\documentclass[a4paper,12pt]{article}
\usepackage[utf8]{inputenc}

\usepackage{graphicx}
\usepackage{amsmath}
\usepackage{epstopdf}
\usepackage[left=2cm,right=1.5cm,
    top=2cm,bottom=2cm,bindingoffset=0cm]{geometry}
\usepackage{cite}

\begin{document}
\title{Nondestructive method of thin film dielectric constant measurements by two-wire capacitor }
\author{ Svitlana Kondovych\textsuperscript{\textsf{1,*}},  Igor Luk'yanchuk\textsuperscript{\textsf{1,2}}}
\date{}
\maketitle
\begin{center}
\textsuperscript{1}\, Laboratory of Condensed Matter Physics, University of Picardie, \\ 33 rue St. Leu, 80039 Amiens, France\\
\textsuperscript{2}\,
ITMO University,  49 Kronverksky Pr., St. Petersburg, 197101
Russia\\
*\, {e-mail:
  \textsf{svitlana.kondovych@gmail.com}}
\end{center}
\renewcommand{\abstractname}{}
 \renewcommand{\figurename}{Fig.}
\abstract{ We suggest the nondestructive method for determination of the dielectric constant of substrate-deposited thin films by capacitance measurement with two parallel wires placed on top of the film. The exact analytical formula for the capacitance of such system is derived. The functional dependence of the capacitance on dielectric constants of the film, substrate and environment media and on the distance between the wires permits to measure the dielectric constant of thin films for the vast set of parameters where previously proposed approximate methods are less efficient.}
\bigskip 

\bigskip  
%%%%%%%%%%%%%%%%%%%%%%%%%%%%%%%%%%%%%%%%%%%%%%%
\begin{figure*}[h!]
	\center
	\includegraphics[width=0.6\linewidth]{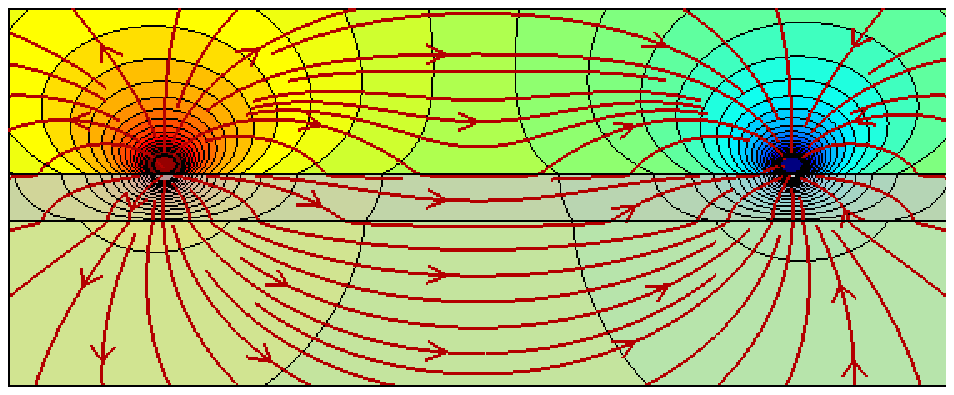}
	\\
	{Electric field of a
two-wire capacitor, \\designated for measurement of the dielectric constant \\of
the substrate-deposited film}
\end{figure*}
%%%%%%%%%%%%%%%%%%%%%%%%%%%%%%%%%%%%%%%%%%%%%%%

\clearpage
\section{Introduction} \
Miniaturization of electronic devices down to the nano-scale has become possible by
achieving the unprecedentedly efficient material functionalities not
available in bulk systems. A large variety of novel nanoscale materials extends from thin films and superlattices \cite{Shi2003book,Lakhtakia2005book,Ramesh2007,Zhang2010book,Hass2013book} to nanoparticles and particle composites (see e.g. \cite{chatzigeorgiou2015interface} and references therein), the unique properties of which open a way to various implementations for nanoelectronics. In particular, tailoring the properties of
substrate-deposited thin films by strain has attracted particular attention
due to technological feasibility and various potential applications such as sensors,
actuators, nonvolatile memories, bio-membranes,  photovoltaic cells,  tunable microwave circuits and micro- and
nano- electromechanical systems  \cite{Shi2003book,Lakhtakia2005book,Ramesh2007,Zhang2010book,Hass2013book}. Control and measurement of
the dielectric constant $\varepsilon$ of thin films present one of the
major objectives of strain-engineering technology to achieve the optimal
dielectric properties of constructed nanodevices.

The arising difficulty, however, is that the conventional technique for
measurement of $\varepsilon $, consisting in the determination of capacitance of
a two-electrode plate capacitor, $C=\varepsilon _{0}\varepsilon S/h$
(where $\varepsilon _{0}$ is the vacuum permittivity, $S$ is the electrode
surface and $h$ is the distance between plates), is not suitable here. The
bottom-electrode deposition at the film-substrate interface, if ever
possible, perturbs the functionality and integrity of the device, whereas
the top-electrode can influence the optical characterization of the system.
In addition, defect-provided leakage currents across thin film can distort the results. The emergent technique of nanoscale capacitance microscopy \cite{Shao2003,Gomila2008} that measures the capacitance between
an atomic force microscope tip and the film is also limited by the same requirement of film deposition on a conductive substrate.

A non-destructive way to overcome these difficulties consists in
employing a capacitor in which both electrodes are located outside but
in close proximity to the film. The capacitance of the system will depend on
its geometry and in particular on the dielectric constants of film and
substrate that finally permits to measure $\varepsilon $. However,
determination of such functional dependence is the complicated electrostatic
problem that, in general, requires cumbersome numerical calculations.
The semi-analytical method of capacitance calculation for a particular case of
planar capacitor in which two semi-infinite electrode plates with parallel,
linearly aligned edges are deposited on the top of the film was proposed by
Vendik \textit{et al.} \cite{Vendik1999}. This geometry attracted the
experimental audience due to the simplicity and intuitive clarity of the
resulting formula. Under the reasonable experimental conditions, the
capacitance of the planar capacitor was found to be inversely proportional
to the width of the edge-separated gap transmission line, $d$, and can be
approximated as $C=\varepsilon _{0}\varepsilon S/d$ where $S$ is the
cross-sectional area of the film below the electrode edge. This expression
is formally equivalent to the capacitance of a parallel-plate capacitor of thickness $d$,
in which the electrodes correspond to the cross-sectional regions.

Note, however, that Vendik's method is limited to the case when the dielectric
constant of the film (we set it as $\varepsilon _{2}$) is much bigger than
the dielectric constants of the environment media, $\varepsilon _{1}$, and the
substrate, $\varepsilon _{3}$, and when the transmission gap is thinner than the film thickness \cite{Deleniv1999}. This restriction is related to the used
``partial capacitance'' or ``magnetic wall'' approximations in which the film,
the substrate and the environment space are assumed to be electrostatically
independent of each other and the electric field lines do not emerge from
the deposited film. Being justified for the upper subspace, which is
normally air with $\varepsilon _{1}=1$, the partial capacitance
approximation can be not accurate enough if the dielectric constant of the substrate is bigger than
(or comparable to) that of the film.

The objective of the present work is to propose the procedure for non-destructive measurements of the dielectric constant of the films,
valid for \textit{any} types of the substrate and environment media. We
consider the geometry in which two parallel wire electrodes are placed on top of the film and derive the \textit{exact} formula for the capacitance of
such system. Our calculations don't imply the partial capacitance
approximation and therefore are valid for nanofilm-substrate devices based on the vast class of materials, extending from
semiconductors to oxide multiferroics.

\section{Model} 
\begin{figure}[t!]
	\center
	\includegraphics[width=0.6\linewidth]{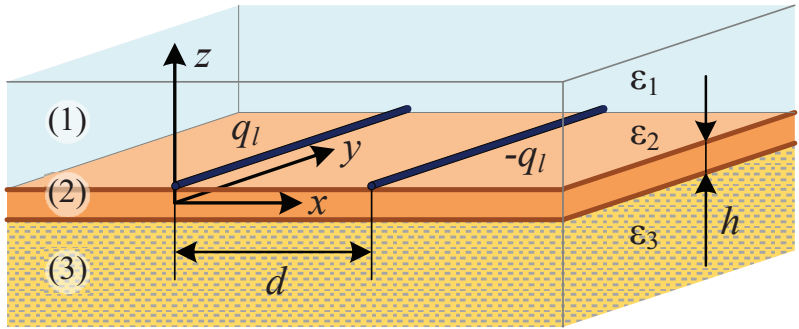}
	\caption{Geometry of the system. Thin film  of thickness $h$ with dielectric constant $\varepsilon_2$ (region (2)) is deposited on the substrate with dielectric constant $\varepsilon_3$ (region (3) at the bottom) and is surrounded by the external environment with dielectric constant $\varepsilon_1$ (region (1) at the top). Two parallel oppositely charged wires with linear charge densities $\pm q_l$ and of radius $R$ (not shown) are placed on top of the film. The distance between wires, $d$, is much larger then $h$ and $R$. Measurement of the capacitance of the two-wire system, $C_l$, (calculated per unit of length) permits to find $\varepsilon_2$. The $z$-axis of the rectangular coordinate system is directed across the film plane, the in-plane $x$-axis is perpendicular to the wires and the $y$-axis is directed along the wires.}
	\label{model}
\end{figure}

\
The geometry of the system is shown in Fig.~\ref{model}. Two parallel wires
with opposite linear charge densities, $\pm q_{l}$, are located on top of the
ferroelectric film. The distance between the
wires, $d$, is much larger than their radius, $R$, and the film thickness, $h$. We also account for anisotropy of the film, assuming that the in-plane
(transverse) dielectric constant differs from the out-of-plane
(longitudinal) one, $\varepsilon _{2}$, by the anisotropy factor $\gamma
^{2}$ and is equal to $\gamma ^{2}\varepsilon _{2}$. The origin of  the rectangular coordinate system is selected in the middle of the film, just below the left wire. The $z$-axis is directed perpendicular to the film plane, the $y$-axis is directed along the wires and the $x$-axis is perpendicular to them. Thus, left and right wires have the coordinates $(0,y,h/2)$ and $(d,y,h/2)$ correspondingly. The translational symmetry of this
system in $y$-direction permits to reduce the consideration to the
2D space, $(x,z)$.
%%%%%%%%%%%%%%%%%%%%%%%%%%%%%%%%%%%%%%%%%%%%%%%
%%%%%%%%%%%%%%%%%%%%%%%%%%%%%%%%%%%%%%%%%%%%%%%

\section{Method} \  %subsection
Using the methods of electrostatics we calculate the distribution of the electrostatic
potential induced by one of the wires (left one). The corresponding Poisson equations have to be written separately for each constituent part of the system (Fig.~\ref{model}), the external 
environment space (region 1), film (2) and substrate (3) :
\begin{equation}
	\begin{array}{ll}
		\partial _{x}^{2}\varphi _{1}+\partial _{z}^{2}\varphi _{1}=-\frac{1}{%
			\varepsilon _{0}\varepsilon _{1}}\rho (x,z), & \quad z>h/2, \\ 
		\gamma^2 \partial _{x}^{2}\varphi _{2}+\partial _{z}^{2}\varphi _{2}=0, & 
		\quad \left\vert z\right\vert <h/2, \\ 
		\partial _{x}^{2}\varphi _{3}+\partial _{z}^{2}\varphi _{3}=0, & \quad
		z<-h/2,%
	\end{array}
	\label{El}
\end{equation}%
where $\rho (x,z)=q_{l}\delta (x)\delta (z-h/2)$ is the charge distribution of the wire.
The electrostatic boundary conditions are applied at the interfaces between
the regions. We set $\varphi _{1}=\varphi _{2}$ and $\varepsilon
_{1}\partial _{z}\varphi _{1}=\varepsilon _{2}\partial _{z}\varphi _{2}$ for
the located at $z=h/2$ environment - film interface, (1)-(2), and $\varphi _{2}=\varphi _{3}$
and $\varepsilon _{2}\partial _{z}\varphi _{2}=\varepsilon _{3}\partial
_{z}\varphi _{3}$ for the located at $z=-h/2$  film - substrate interface, (2)-(3).

%%%%%%%%%%%%%%%%%%%%%%%%%%%%%%%%%%%%%%%%%%%%%%%
\begin{figure}[t!]
\begin{center}
\begin{minipage}[h]{0.482\linewidth}
\includegraphics*[width=1\linewidth]{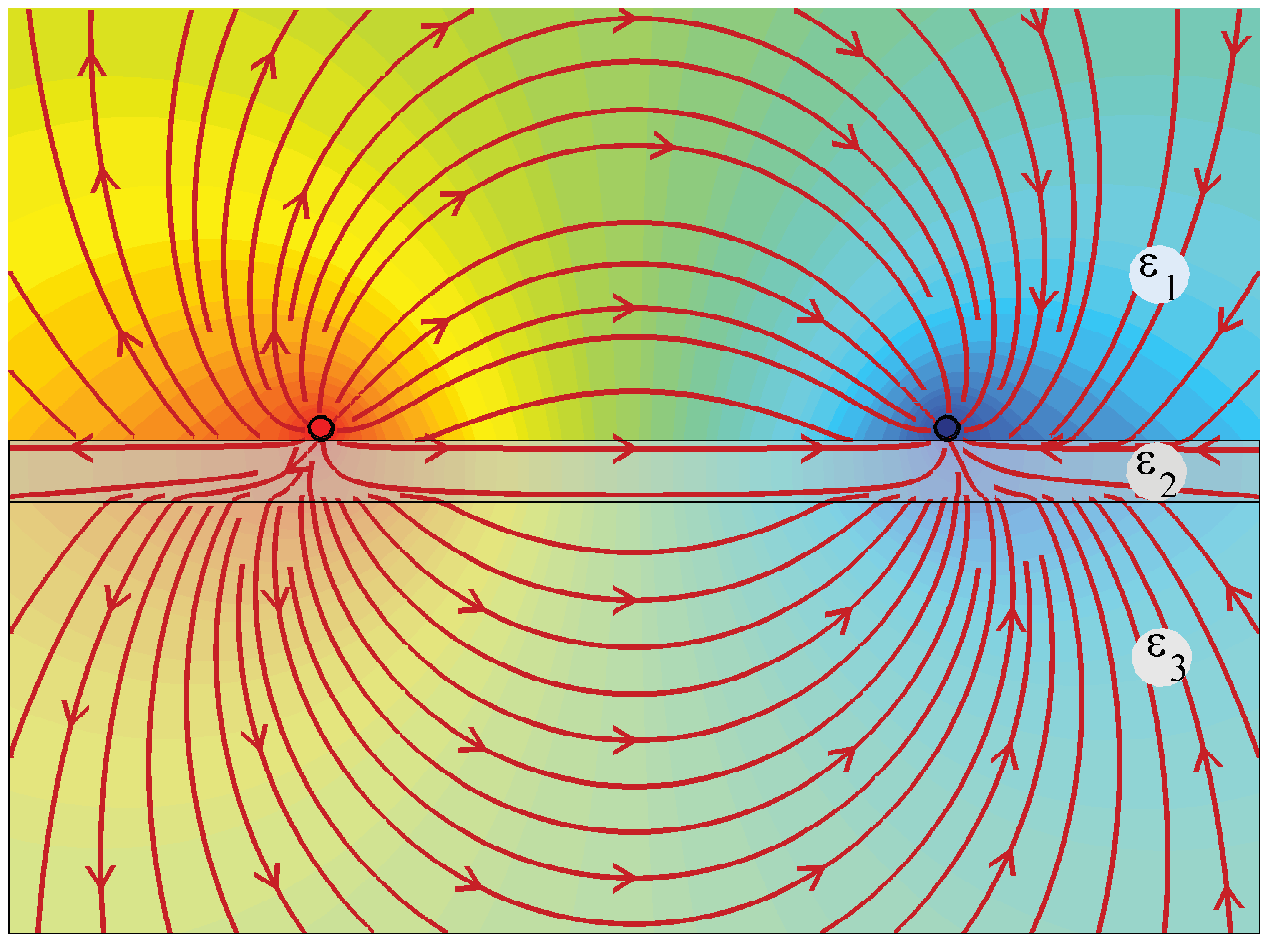} 
\end{minipage}
\hfill 
\begin{minipage}[h]{0.476\linewidth}
\includegraphics*[width=1\linewidth]{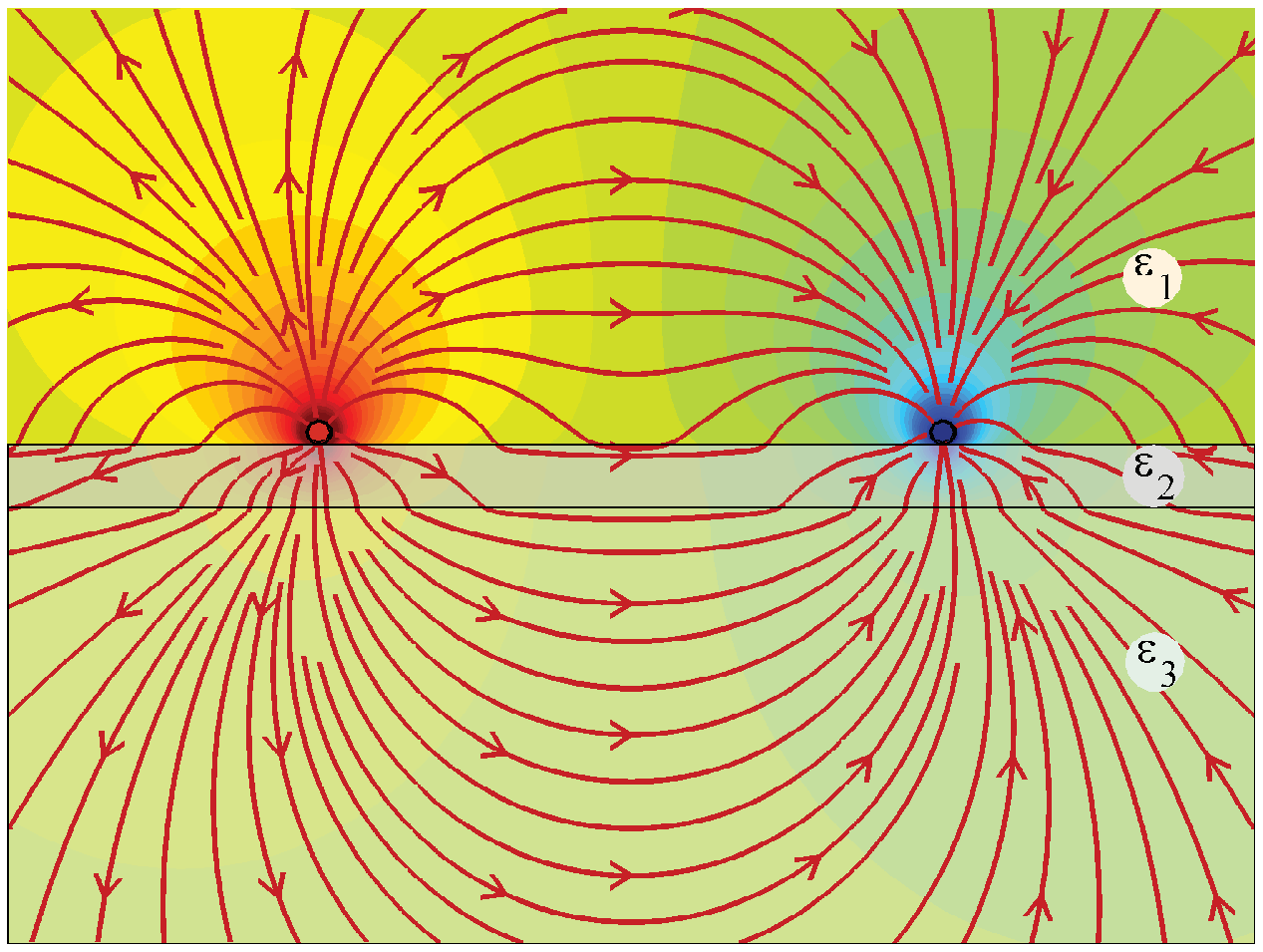}
\end{minipage}
\begin{minipage}[h]{1\linewidth}
\begin{tabular}{p{0.47\linewidth}p{0.47\linewidth}}
\centering a) $\varepsilon _{2}\geq \varepsilon _{3}\gg \varepsilon _{1}$ & \centering b) $\varepsilon _{3}\geq \varepsilon _{2}\gg \varepsilon _{1}$ \\
\end{tabular}
\end{minipage}
\vspace*{-.5cm}
	\caption{%
		Electric field lines and corresponding electrostatic potential (colour map) induced by two oppositely charged parallel wires located on top of the substrate-deposited film and directed perpendicular to the figure plane. The geometry of the system is depicted in Fig.~\ref{model}. a) For the high-$\varepsilon$ film with  $\protect\varepsilon _{2}\geq 
		\protect\varepsilon _{3}\gg \protect\varepsilon _{1}$. b) For the low-$\varepsilon$ film with  $\protect\varepsilon _{3}\geq 
		\protect\varepsilon _{2}\gg \protect\varepsilon _{1}$}
	\label{lines}
\end{center}
\end{figure}
%%%%%%%%%%%%%%%%%%%%%%%%%%%%%%%%%%%%%%%%%%%%%%%

The Fourier method, similar to the one applied in \cite{Rytova1967,Baturina2013} for point charges, is used to solve the system (\ref{El}) and find the relevant asymptotes. Following this method, the
cos-Fourier transform of the potential inside the film, $\widetilde{\varphi }%
_{2}(k,z)={\int\limits_{0}^{\infty }\varphi _{2}(x,z)\cos }\left( kx\right) {%
	dx}$, is found as: 
\begin{equation}
	\widetilde{\varphi }_{2}=\frac{q_{l}e^{\gamma kh/2}}{2\varepsilon
		_{0}k\left( \varepsilon _{2}+\varepsilon _{1}\right) }\frac{\frac{%
			\varepsilon _{2}-\varepsilon _{3}}{\varepsilon _{2}+\varepsilon _{3}}%
		e^{-kz}+e^{\gamma kh}e^{kz}}{e^{2\gamma kh}-1+2\gamma \frac{h}{\lambda }}.
	\label{Pot}
\end{equation}%
Here, $\lambda $ is a characteristic length of the system, 
\begin{equation}
	\lambda =\gamma \frac{\left( \varepsilon _{2}+\varepsilon _{1}\right) \left(
		\varepsilon _{2}+\varepsilon _{3}\right) }{\varepsilon _{2}\left(
		\varepsilon _{1}+\varepsilon _{3}\right) }h,  \label{lambda}
\end{equation}%
that will be used below to delimit the regions with a different spatial decay of ${\varphi _{2}}$
in the $x$-direction.
The inverse transformation of Eq.~(\ref{Pot}), 
\begin{equation}
	{\varphi _{2}}={\frac{2}{\pi }\int\limits_{0}^{\infty }\widetilde{\varphi }%
		_{2}\cos }\left( kx\right) {dk,}  \label{potential}
\end{equation}%
permits to find the expression for ${\varphi _{2}(x,z)%
}$. Similar calculations can be done for ${\varphi _{1}(x,z)}$ and  ${\varphi _{3}(x,z)}$. The results of the numerical solution of Eqs.~(\ref{El}) for two typical sets of dielectric constants $\varepsilon_1$, $\varepsilon_2$ and $\varepsilon_3$ are presented in Fig.~\ref{lines}.

\section{Results} 
\begin{figure}[h!]
	\center
	\includegraphics[width=0.4\linewidth]{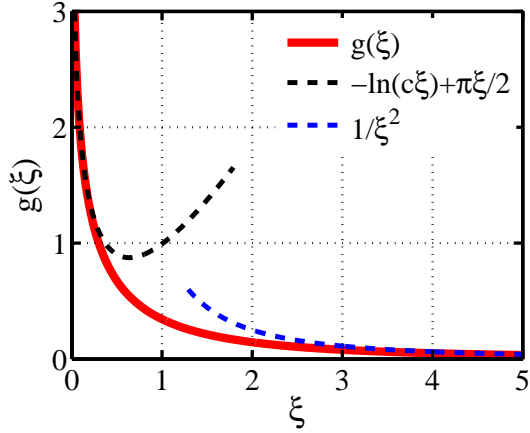}
	\caption{Auxiliary function $g(\xi)=\left( \frac{\pi }{2}-\mathrm{Si\,}\xi\right) \sin \xi-\mathrm{%
			Ci\,}\xi\cos \xi $ and its small-$\xi$ and large-$\xi$ asymptotes. $\ln c\simeq 0.577$ is the Euler's constant.}
	\label{g_func}
\end{figure}

\ %subsection
Having calculated the potential induced by one of the wires and taking into
account their equivalence we can find the capacitance of the system per
unit of length as $C_{l}=q_{l}/\Delta \varphi $ where $\Delta \varphi ={%
	\varphi _{2}(R,h/2)-\varphi _{2}(d-R,h/2)}$ is the potential difference
between the wires. For the large wire separation, $d\gg h,R$, the first term in $%
\Delta \varphi $ contributes as the $d$-independent cutoff constant, whereas
the second one can be calculated analytically, by an expansion of (\ref{Pot})
in series over the small parameter $kh\ll 1$ that allows for exact
integration in Eq.~(\ref{potential}). Finally, we obtain the following
expression for the inverse capacitance, 
\begin{equation}
	C_{l}^{-1}={\frac{\left(\pi\varepsilon_0 \right)^{-1}}{ \varepsilon
			_{1}+\varepsilon _{3} }}\left[ \ln A\frac{d}{\lambda }+\left( 1-\frac{%
		h}{\lambda }\beta \right) g\left( \frac{d}{\lambda }\right) \right] ,
	\label{capacitance}
\end{equation}
where $A$ is the non-essential for further analysis constant that comprises
the wire-scale cut-off, 
\begin{equation}
	\beta =\gamma +\frac{{\varepsilon _{3}}}{2\varepsilon _{2}}{\left( 1+\gamma
		\right) }  \label{beta}
\end{equation}
and 
\begin{equation}
	g\left( x\right) =\left( \frac{\pi }{2}-\mathrm{Si\,}x\right) \sin x-\mathrm{%
		Ci\,}x\cos x  \label{gfunct}
\end{equation}%
is the shown in Fig.~\ref{g_func} auxiliary function composed from the Sine and Cosine
Integrals \cite{Abramowitz1988book}.

%%%%%%%%%%%%%%%%%%%%%%%%%%%%%%%%%%%%%%%%%%%%%%%
%%%%%%%%%%%%%%%%%%%%%%%%%%%%%%%%%%%%%%%%%%%%%%%
\bigskip

\begin{figure}[ht!]
	\center
	\begin{minipage}[h]{0.48\linewidth}
\includegraphics*[width=1\linewidth]{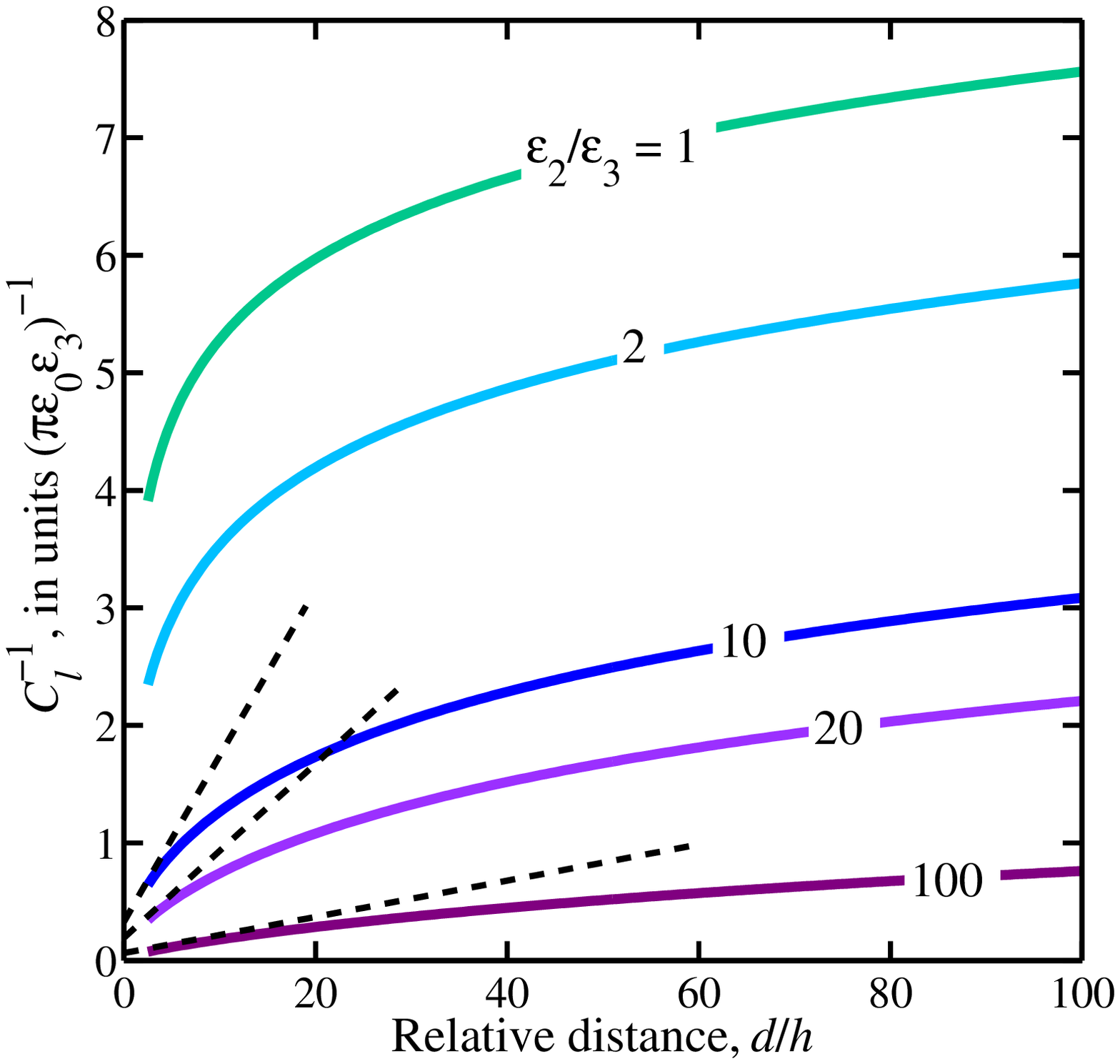} 
\end{minipage}
\hfill 
\begin{minipage}[h]{0.48\linewidth}
\includegraphics*[width=1\linewidth]{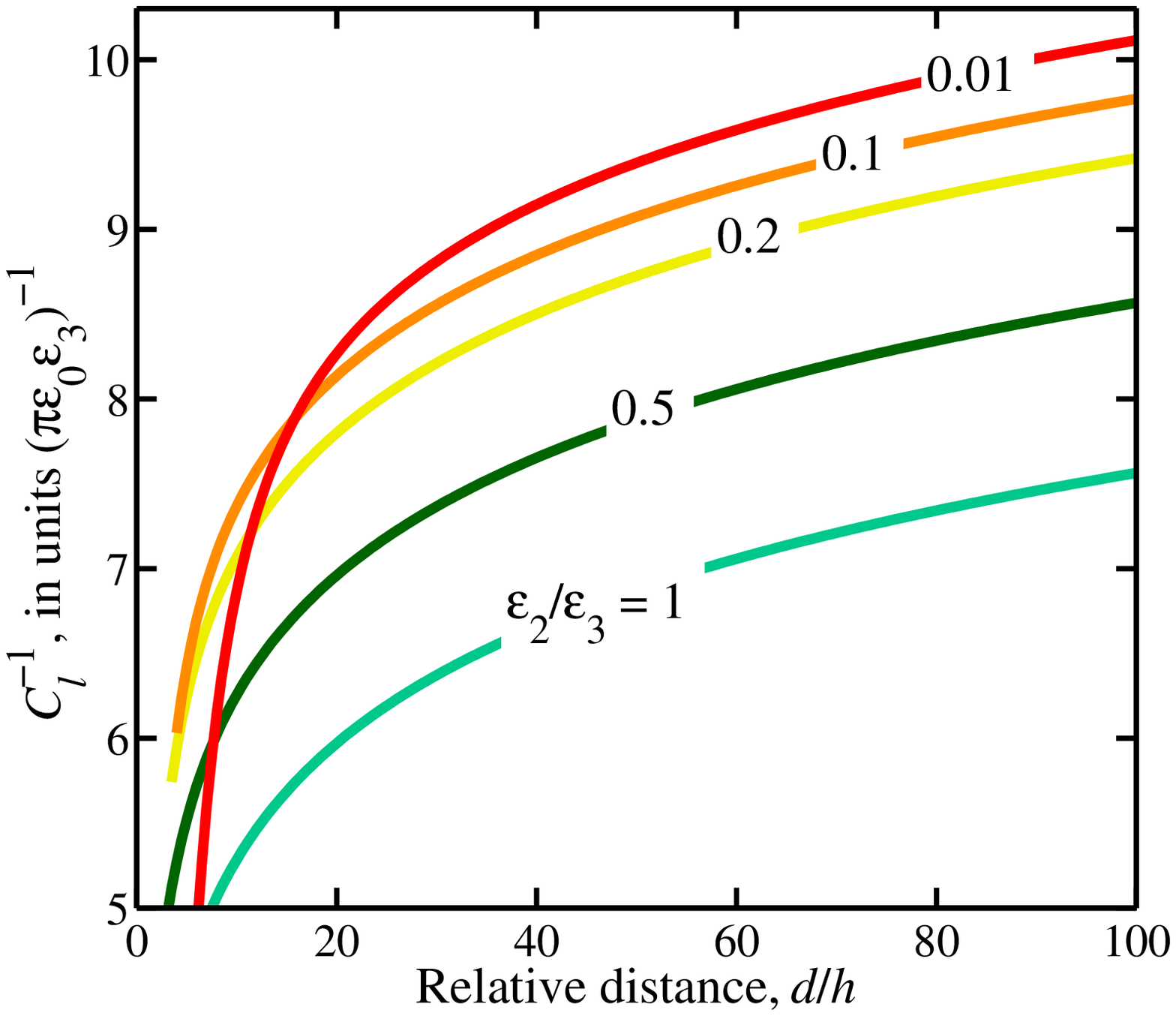}
\end{minipage}
\begin{minipage}[h]{1\linewidth}
\begin{tabular}{p{0.47\linewidth}p{0.47\linewidth}}
\centering a) & \centering b) \\
\end{tabular}
\end{minipage}
\vspace*{-.5cm}
	\caption{The inverse capacitance of the system, $C^{-1}_l$, in  units  $(\pi\varepsilon_0\varepsilon_3)^{-1}$, as a function of the relative distance between the wires, $d/h$, for different ratios of the film and substrate dielectric constants, $\varepsilon_2/\varepsilon_3$. The dielectric constant of the environment media is assumed to be small, $\varepsilon_1\simeq 1$. a) For the high-$\varepsilon$ film with  $\protect\varepsilon _{2}\geq 
		\protect\varepsilon _{3}\gg \protect\varepsilon _{1}$. b) For the low-$\varepsilon$ film with  $\protect\varepsilon _{3}\geq 
		\protect\varepsilon _{2}\gg \protect\varepsilon _{1}$.}
	\label{fig_cap}
\end{figure}

Given by Eq.~(\ref{capacitance}) dependence of the system capacitance
on the distance between the wires presents the basic result for
determination of the dielectric constant of the film that enters there
through two fitting parameters, $\lambda (\varepsilon _{2})$ and $\beta
(\varepsilon _{2})$. We discuss now in detail how this procedure can be
implemented in practice, considering for simplicity the isotropic film with $%
\gamma =1$, encompassed by the external environment with $\varepsilon _{1}\ll
\varepsilon _{2},\varepsilon _{3}$ that gives $\beta =1+{\varepsilon
	_{3}/\varepsilon _{2}}$ and $\lambda =\left( 1+\varepsilon _{2}/\varepsilon
_{3}\right) h$. We analyze separately the cases of high-$\varepsilon $ and
low-$\varepsilon $ films (with $\varepsilon _{2}\geq \varepsilon _{3}\gg
\varepsilon _{1}$ and $\varepsilon _{3}\geq \varepsilon _{2}\gg \varepsilon
_{1}$ correspondingly) that have different electrostatic behavior. As
shown in Fig.~\ref{lines}, wires-induced electric field lines are ``repelled'' from the film in the
first case (Fig.~\ref{lines},a) and ``captured'' by the film in the second one (Fig.~\ref{lines},b). Fig.~\ref{fig_cap} presents given by  Eq.~(\ref{Pot}) dependence of the inverse capacitance $C_{l}^{-1}$, measured in units  $(\pi\varepsilon_0\varepsilon_3)^{-1}$ $=3.6\times 10^4 \varepsilon_3^{-1} \mu$m/pF, on the relative distance between the wires, $d/h$, for both cases.

%%%%%%%%%%%%%%%%%%%%%%%%%%%%%%%%%%%%%%%%%%%%%%%
%%%%%%%%%%%%%%%%%%%%%%%%%%%%%%%%%%%%%%%%%%%%%%%

\paragraph{High-$\protect\varepsilon $ film, $\protect\varepsilon _{2}\geq 
	\protect\varepsilon _{3}\gg \protect\varepsilon _{1}$.}

For a large ratio $\varepsilon _{2}/\varepsilon _{3}$ the characteristic scale $%
\lambda $ can be comparable and even larger than the linear size of the
system, and therefore the $g$-function can be expanded over the small
parameter $d/\lambda $ as $g\left( d/\lambda \right) \simeq -\ln \left(
cd/\lambda \right) +\pi d/2\lambda $ \cite{Abramowitz1988book}, (Fig.~\ref{g_func}), where $\ln c$ is the Euler's constant, $\ln c \overset{%
	n\rightarrow \infty }{=}\Sigma _{k=1}^{n}\,k^{-1}-\ln n\simeq 0.577$. Then, the resulting
expression for $C_{l}^{-1}$ can be simplified to: 
\begin{equation}
	C_{l}^{-1}=\mathrm{const}+{\frac{1}{\varepsilon _{0}\varepsilon _{2}}}\frac{d%
	}{2h},  \label{Chigh}
\end{equation}%
that permits to measure $\varepsilon _{2}$ via the linear slope of
dependence $C_{l}^{-1}(d)$ at $d\rightarrow 0$ (Fig.~\ref{fig_cap},a). This method is
analogous to that for geometry of planar capacitor with semi-infinite plates
\cite{Vendik1999} due to the similar linear dependence on the distance between
electrodes. Presented in Fig.~\ref{fig_cap},a numerical analysis shows, however, some
restrictions for the application of this method. The distance between electrodes
at which the linearity is manifested should be rather small (but still
larger than $R$ and $h$) and the parameter ${\varepsilon _{2}/\varepsilon
	_{3}}$ should be large enough.

\paragraph{Low-$\protect\varepsilon $ film, $\protect\varepsilon _{3}\geq 
	\protect\varepsilon _{2}\gg \protect\varepsilon _{1}$. }
For small $\varepsilon _{2}/\varepsilon _{3}$ the opposite situation, $%
d>\lambda $, takes place and the large-scale approximation for the $g$-function
can be used, $g\left( d/\lambda \right) \simeq \left( \lambda /d\right) ^{2}$ \cite{Abramowitz1988book}. Then, Eq.~(\ref{capacitance}) is simplified to:
\begin{equation}
	C_{l}^{-1}\simeq {\frac{1}{\pi \varepsilon _{0}\varepsilon _{3}}}\left[ \ln A%
	\frac{d}{h}-\frac{\varepsilon _{3}}{\varepsilon _{2}}\frac{h^{2}}{d^{2}}%
	\right] ,  \label{Csmall}
\end{equation}%
the corresponding dependencies $C_{l}^{-1}(d/h)$ being shown in Fig.~\ref{fig_cap},b.

%%%%%%%%%%%%%%%%%%%%%%%%%%%%%%%%%%%%%%%%%%%%%%%
\begin{figure}[b!]
	\center
	\includegraphics[width=0.6\linewidth]{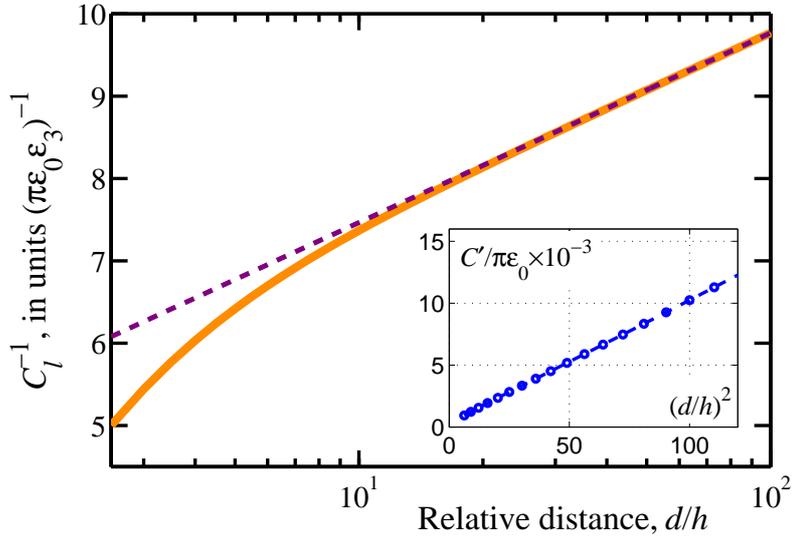}
	\caption{Determination of the dielectric constant of the low-$\varepsilon$ film. The inverse capacitance of the system, $C^{-1}_l$,  is measured in  units $(\pi\varepsilon_0\varepsilon_3)^{-1}$ and is plotted as a function of the logarithm of the relative distance, $d/h$, between the wires (orange solid line). Then, it is extracted from the linear background (purple dashed line), determined from the slope of $C^{-1}_l$ at $d\rightarrow\infty$. The resulting capacitance, $C^\prime$, is plotted in units  $\pi\varepsilon_0$ as a function of $(d/h)^2$, giving the straight line (inset). The tangent coefficient corresponds to the film dielectric constant $\varepsilon_2$ (here, $\varepsilon_2=100$).}
	\label{method}
\end{figure}
%%%%%%%%%%%%%%%%%%%%%%%%%%%%%%%%%%%%%%%%%%%%%%%

To extract the value of $\varepsilon _{2}$ from experimental data one should
first get rid of the  $\varepsilon _{2}$-independent contribution presented
by the logarithmic term in Eq.~(\ref{Csmall}), which contains the unknown cut-off
constant. For this, one can plot $C_{l}^{-1}$ in units $(\pi \varepsilon _{0}\varepsilon
_{3})^{-1}$ vs. $\ln (d/h)$ as shown in Fig.~\ref{method} and subtract the linear
background, manifested at $d\rightarrow \infty $. The residual
contribution to the capacitance, $C_{l}^{\prime }=$ $C_{l}^{\infty }-C_{l}$, is
given by the simple dependence $C_{l}^{\prime }=\pi \varepsilon _{0}\varepsilon
_{2}(d/h)^2$, independent of the value of $\varepsilon_3$. Then, the dielectric constant, $\varepsilon _{2}$,  can
be extracted from the slope of $C_{l}^{\prime }$, plotted in units $\pi \varepsilon _{0}$ as a function of 
$%
(d/h)^{2}$ (Inset to Fig.~\ref{method}). 

Note that for the low-$\varepsilon$ films the small ratio $d/\lambda<1$ can be realized only for the distances $d$ much smaller than the cutoff lengths, $R$ and $h$. Therefore the linear approximation over $d/\lambda$, used in \cite{Vendik1999}, makes no sense here.

\section{Conclusion} \ 
The explicit analytical expression (\ref{capacitance}) derived  for the capacitance  of two parallel wires placed on top of the substrate-deposited film gives a way for experimental non-destructive measurements of the dielectric constant of this film. For experimental implementation, it can be convenient to deposit the system of equidistant wires and measure consequently the capacitance between them. The technical procedure consists in the determination of the capacitance as a function of the distance between the wires with subsequent comparison (fit) with functional dependence, given by Eq.~(\ref{capacitance}). Simple and intuitively clear realizations of this method for high-$\varepsilon$ and low-$\varepsilon$ films (with respect to substrate) are proposed. The suggested procedure is based on the exact expression that permits to measure the dielectric constant for those systems in which traditionally used techniques are less precise or even fail because of the uncontrolled approximations.

\bigskip

	We acknowledge the stimulating discussions with T. Baturina, V. Vinokur and A. Razumnaya.   
	This work was supported by FP7-ITN-NOTEDEV project.

\providecommand{\othercit}{}
\providecommand{\jr}[1]{#1}
\providecommand{\etal}{~et~al.}

\end{document}